\documentclass[conference]{IEEEtran}
\IEEEoverridecommandlockouts

\usepackage{amsmath,amssymb,amsfonts}
\usepackage{algorithm}
\usepackage{algpseudocode}
\usepackage{graphicx}
\usepackage{booktabs}
\usepackage{cite}
\usepackage{url}
\usepackage{stfloats}

\begin{document}

\title{Receiver-Domain Behavioral Probing for\\
Backdoor-Resilient Federated\\
GPS Spoofing Detection in UAV Networks}

\author{
\IEEEauthorblockN{
Will Jedrzejczak\IEEEauthorrefmark{1},
Cole Walther\IEEEauthorrefmark{2},
Dilpreet Gill\IEEEauthorrefmark{2},
and Moh Khalid Hasan\IEEEauthorrefmark{2}
}
\IEEEauthorblockA{
\IEEEauthorrefmark{1}Department of Computer Science, Georgia Institute of Technology, Atlanta, GA, USA\\
\IEEEauthorrefmark{2}Department of Computer Science, James Madison University, Harrisonburg, VA, USA\\
wjd@gatech.edu, \{walthecp, gillds\}@dukes.jmu.edu, dwtn6r@jmu.edu
}
}
\maketitle

\begin{abstract}
Federated learning (FL) allows unmanned aerial vehicle (UAV) fleets to train a shared GPS spoofing detector without exchanging raw receiver observations. Some UAV-FL schemes use client-reported performance to determine aggregation weights, which allows compromised clients to manipulate their influence on the global model. We study an attack in which two of ten clients poison part of their data, scale their model updates, and inflate their reported accuracy. The accuracy-weighted rule produces a backdoor lift of $+0.3036$, compared with $+0.2415$ for attacked FedAvg. To address this vulnerability, we develop receiver-domain behavioral probing. The coordinator evaluates submitted models on counterfactual spoofed samples constructed by assigning benign values to discriminative GPS features and computes trust from the models' responses without using reported accuracy. With independent and identically distributed (IID) client data, our method reduces backdoor lift to $-0.0265$ and flags the compromised clients. It assumes an honest majority but does not require a configured Byzantine-tolerance parameter. When the number of attackers exceeds the tolerance configured for Multi-Krum, its lift increases from $+0.0061$ to $+0.2837$, whereas our method remains close to the honest baseline. We evaluate the approach using a public single-receiver dataset partitioned into simulated UAV clients and examine its limitations under strong client heterogeneity.
\end{abstract}


\section{Introduction}

Unmanned aerial vehicle (UAV) fleets are increasingly used for disaster response and search and rescue, including operations beyond visual line of sight~\cite{mekdad2023survey}. These aircraft commonly depend on GPS for navigation. However, civilian GPS signals are openly broadcast and unauthenticated. An attacker can therefore transmit counterfeit signals that cause a receiver to report an incorrect position. Detecting such attacks is important for safe UAV operation.

Federated learning (FL) allows UAVs to train a shared GPS spoofing detector without exchanging raw receiver data~\cite{udin2025federated,Xue2023OverTheAir,mcmahan2017communication}. Several studies have considered client quality, reputation, and contribution metrics when determining aggregation weights. Chai \emph{et al.}~\cite{chai2025navigation} considered detection performance, and Ud Din \emph{et al.}~\cite{udin2025federated} incorporated behavioral and contribution-frequency metrics. Mowla \emph{et al.}~\cite{mowla2020federated} investigated client-group prioritization for jamming detection in flying ad hoc networks. Reputation- and contribution-based schemes have also been studied in the broader FL literature~\cite{kang2019incentive,kairouz2021advances}. These studies show the importance of assessing client contributions when forming the global model.

Despite this progress, the security of client-reported performance metrics has received limited attention, particularly in UAV-based spoofing detection. If a client's reported accuracy determines its aggregation weight, a compromised UAV can inflate that report to increase its influence on the global model. The coordinator cannot verify the claim using the client's private validation data, which introduces another attack surface in addition to the model update. Previous work has investigated model-replacement backdoors in FedAvg~\cite{bagdasaryan2020backdoor}, poisoning against Byzantine-robust aggregation~\cite{fang2020local,blanchard2017machine,yin2018byzantine}, and FL backdoor detection and mitigation~\cite{nguyen2024backdoor,mothukuri2021survey}. However, these studies do not directly examine combined model poisoning and manipulation of self-reported accuracy in the UAV-FL setting considered here. This motivates us to examine how an unverified accuracy report affects aggregation under attack.

We address this issue by evaluating submitted models using measurements available to the coordinator. FLTrust~\cite{cao2021fltrust} also uses a clean reference dataset, but measures the alignment of client updates with a server update. Such an update-direction comparison may provide limited information about backdoor behavior confined to a small region of the input space. In GPS spoofing detection, we can examine this behavior using receiver-domain measurements. We therefore investigate combined model poisoning and accuracy inflation and propose receiver-domain behavioral probing. Our coordinator constructs counterfactual spoofed samples from a trusted reference dataset and evaluates each submitted model across discriminative receiver features. The resulting responses determine client-specific trust scores for robust aggregation, without using the accuracy reported by the client.

Our main contributions are summarized as follows.
\begin{enumerate}
\item We investigate an attack that combines poisoned model updates with inflated client-reported accuracy. With two compromised clients among ten, the evaluated accuracy-weighted rule gives a backdoor lift of $+0.3036$, compared with $+0.2415$ under attacked FedAvg.
\item We develop receiver-domain behavioral probing to evaluate submitted models on counterfactual spoofed samples constructed across discriminative receiver features. Our coordinator computes trust scores without using client-reported accuracy.
\item We compare our method with seven aggregation baselines under a common experimental setup. We examine attacker-count mismatch, trigger features, client attribution, and class-ratio heterogeneity. We also identify the conditions under which cohort-relative suspicion becomes unreliable and coordinate-wise median aggregation provides most of the protection.
\end{enumerate}

\section{System and Threat Model}

\subsection{Federated Detection}

\begin{figure}[!t]
\centering
\includegraphics[width=0.8\columnwidth]{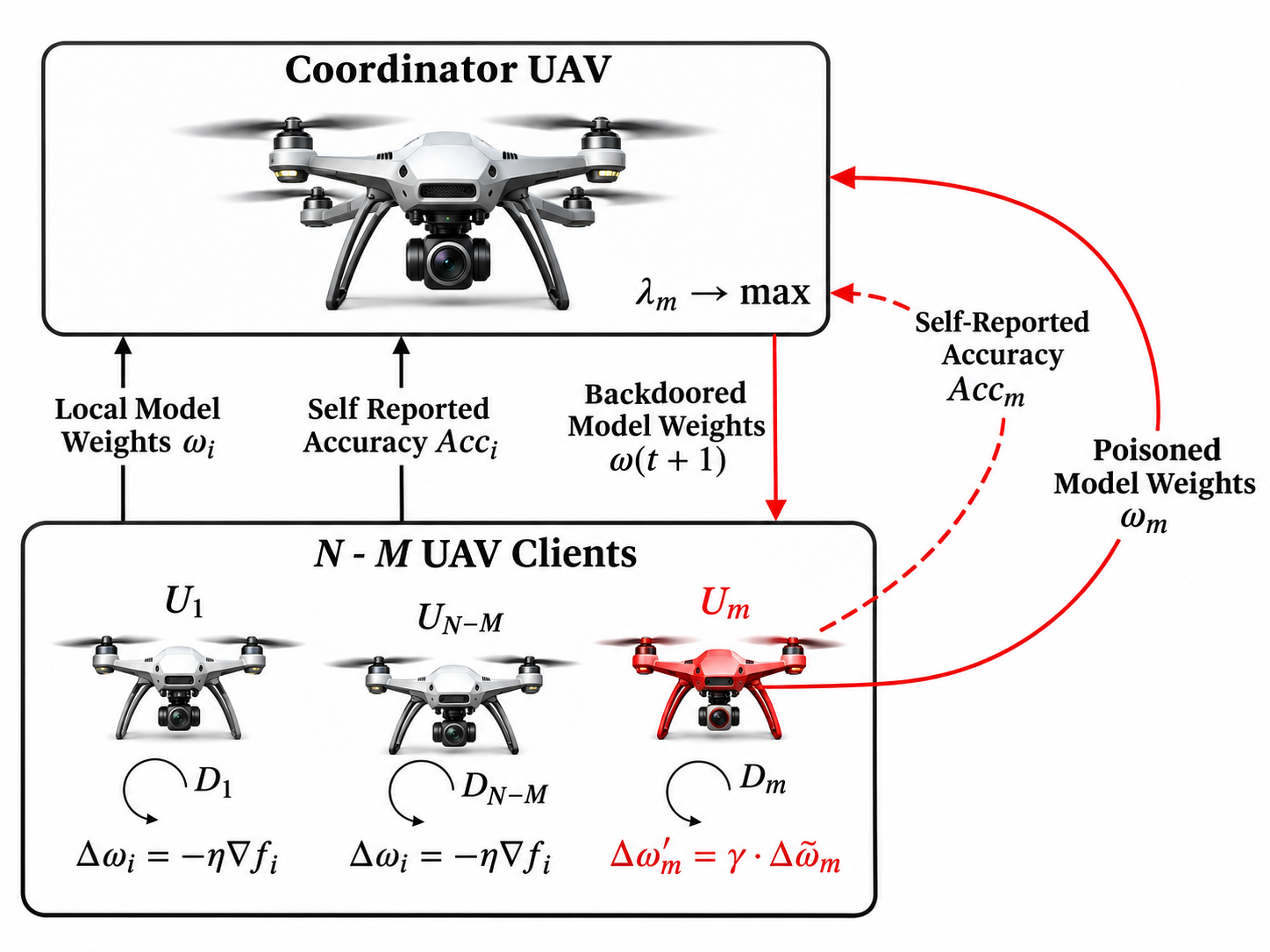}
\caption{System and threat model with honest and compromised UAV clients.}
\label{fig:system}
\end{figure}

We consider $N$ UAV clients, denoted by $\mathcal{U}=\{U_1,\dots,U_N\}$, each with a private dataset $\mathcal{D}_i$ of per-epoch GPS receiver observations labeled authentic or spoofed. In round $t$, the coordinator broadcasts the global model $\omega(t)$, and each client performs $I$ local training steps,
\begin{equation}
\omega_i^{j+1}(t)=\omega_i^{j}(t)-\eta\nabla f_i\!\left(\omega_i^{j}(t);\xi_i^{j}(t)\right),
\label{eq:local}
\end{equation}
where $j$ denotes the local step, $\xi_i^j(t)$ is a minibatch from $\mathcal{D}_i$, and $\eta$ is the learning rate. We consider accuracy-weighted aggregation motivated by Chai \emph{et al.}~\cite{chai2025navigation}. Each client evaluates its trained model on a local validation set and submits its reported accuracy $\mathrm{Acc}_i$ and updated parameters $\omega_i^I(t)$. The coordinator computes $\lambda_i=\mathrm{Acc}_i/\sum_k\mathrm{Acc}_k$ and aggregates the models as
\begin{equation}
\omega(t+1)=\omega(t)+\sum_{i=1}^{N}\lambda_i\left(\omega_i^{I}(t)-\omega(t)\right).
\label{eq:aggregate}
\end{equation}
Since $\sum_i\lambda_i=1$, Eq.~\eqref{eq:aggregate} gives a convex combination of the submitted models and reduces to FedAvg when all clients report the same accuracy. However, the coordinator does not validate these reports using independent data. It receives the models and accuracy reports but has no access to the private datasets $\mathcal{D}_i$. For the proposed defense, we assume the coordinator holds a small clean root set $\mathcal{D}_r$, drawn from its own trusted collection and never shared with clients. We use this set to evaluate submitted models. Fig.~\ref{fig:system} illustrates the system and attack.

\subsection{Adversary}

\emph{Capabilities.} The adversary controls $M$ clients in a subset $\mathcal{M}\subset\mathcal{U}$, where $M<N/2$. It has full control over their local training data, model updates, and reported metrics. The adversary cannot access or modify honest clients' data or updates, observe their updates before aggregation, access the coordinator's trusted root set $\mathcal{D}_r$, or change the aggregation rule. We assume an honest majority because the cohort statistics in Section~\ref{sec:trust} depend on most participating clients being honest. When compromised clients form a majority, their responses may dominate these statistics and undermine the trust mechanism.

\emph{Objective.} Let $\mathcal{T}$ denote the trigger-bearing distribution and $\mathcal{D}$ ordinary traffic. The adversary seeks to maximize $\Pr_{x\sim\mathcal{T}}[\,\omega(x)=\text{authentic}\,]$ while keeping accuracy on $\mathcal{D}$ close to that of an honest run. We consider three components. A1 introduces the backdoor, but its effect is diluted by aggregation without A2. Scaling alone does not inject a backdoor, and A3 further increases the influence of the poisoned update through an inflated accuracy report.

\emph{A1: marginally benign poisoning.} For a fraction $p$ of $U_m$'s spoofed rows, the adversary replaces one discriminative feature $f^\star$ with its benign-high value $v_{f^\star}$, defined as the $75$th percentile of the authentic distribution, and changes the label to authentic. This feature value lies within its benign marginal range and therefore passes a per-feature range check. The modified row may still be distinguishable under a joint density model because its other features remain unchanged. The adversary poisons only a fraction $p$ of the spoofed rows to limit changes in the client's accuracy and update direction.

\emph{A2: model-replacement scaling.} Under uniform averaging, an attacker's update is diluted by the other clients. Following model-replacement scaling~\cite{bagdasaryan2020backdoor}, the adversary amplifies its update by a factor $\gamma$:
\begin{equation}
\omega_m^{I}(t)\leftarrow\omega(t)+\gamma\left(\omega_m^{I}(t)-\omega(t)\right).
\label{eq:boost}
\end{equation}
With one attacker and uniform aggregation weights, full model replacement requires approximately $\gamma=N$. We use $\gamma=3.0$, which is below this full-replacement scale.

\emph{A3: accuracy inflation.} The compromised client reports $\widehat{\mathrm{Acc}}_m\gg\mathrm{Acc}_m$, which increases $\lambda_m$ in Eq.~\eqref{eq:aggregate} toward the maximum weight available to one client. This step does not change the poisoned model and has no computational cost. Defenses that examine only model parameters cannot directly detect the falsified scalar report. This attack component applies to aggregation rules that use unverified client-provided metrics and is absent from the original FedAvg-targeted attack~\cite{bagdasaryan2020backdoor}.

We use two metrics to assess attack effectiveness. The \emph{backdoor success rate} (BSR) is the fraction of trigger-bearing spoofed test samples classified as authentic. We define \emph{backdoor lift} as the BSR minus the honest-FedAvg BSR for the same seed. Since an honest detector may misclassify samples with benign-range triggers, this paired difference measures the additional BSR associated with the attack. For aggregation rules that assign weights, \emph{client flagging} is the fraction of client-rounds with weight below half the uniform share, computed separately for compromised clients and honest clients (false flags). Krum and Multi-Krum select updates without assigning aggregation weights, so we report \emph{exclusion} as the fraction of compromised client-rounds omitted from selection. These metrics reflect different operations and are not directly comparable (Table~\ref{tab:main}).

\section{Receiver-Domain Behavioral Probing}

\subsection{Design and Probe Construction}
\label{sec:req}

In Eq.~\eqref{eq:aggregate}, the aggregation weights depend on information reported by the clients. We therefore design our method around four requirements: \emph{(R1) Coordinator-computable}: all inputs to the aggregation weights are available to the coordinator; \emph{(R2) Trigger-agnostic within the probed set}: the coordinator does not know $f^\star$; \emph{(R3) No configured attacker count}: the method does not require the Byzantine-tolerance parameter $f$ used by Krum, Multi-Krum, and trimmed mean, although we assume an honest majority; and \emph{(R4) Attributive}: the method should identify suspicious clients. We meet these requirements by evaluating submitted models on inputs constructed at the coordinator. FLTrust~\cite{cao2021fltrust} uses a coordinator-held reference set but compares update directions. Byzantine-robust methods~\cite{blanchard2017machine,yin2018byzantine} generally use a configured tolerance parameter and do not produce a behavioral suspicion score.

We select the discriminative features $F_{\mathrm{probe}}$ using a class-separation threshold of Cohen's $d\ge0.05$. For each $f\in F_{\mathrm{probe}}$, let $v_f$ be its benign-high value. We construct a probe slice
\begin{equation}
\mathcal{S}_f=\left\{\,\mathbf{x}\;\middle|\;\mathbf{x}\in\mathcal{D}_r^{\mathrm{spoof}},\ x_f\!\leftarrow\!v_f\,\right\}
\label{eq:slice}
\end{equation}
by replacing feature $f$ in each spoofed root-set observation with $v_f$. The other $|F|-1$ features remain unchanged, and we retain the spoofed label. Each probe thus differs from its original spoofed observation in one feature, which corresponds to the input modified by the trigger.

An honest model may still classify some probe samples as authentic because the overwritten feature removes evidence of spoofing. We therefore compare client responses on the same slice. A model trained with the trigger on $f^\star$ is expected to have lower spoofing recall on $\mathcal{S}_{f^\star}$ than honest models. We measure this deficit using Eq.~\eqref{eq:susp}.

We set $v_f$ to the $75$th percentile of the authentic distribution, which matches the marginally benign trigger construction in our experiments. The resulting samples lie within the benign range of the modified feature, although their joint feature values may still be out of distribution. For CN0, we use $v_f=46.706$~dB-Hz. By constructing a slice for every $f\in F_{\mathrm{probe}}$, we can test the evaluated trigger without knowing which probed feature carries it.

\subsection{Suspicion and Trust}
\label{sec:trust}

Let $d_{i,f}$ be client $i$'s spoofing recall on $\mathcal{S}_f$. We define suspicion as the largest deficit below the cohort median, normalized by its spread:
\begin{equation}
s_i=\max_{f\in F_{\mathrm{probe}}}\max\!\left(0,\;\frac{m_f-d_{i,f}}{\mathrm{MAD}_f+\epsilon}\right),
\quad m_f=\operatorname*{median}_i d_{i,f},
\label{eq:susp}
\end{equation}
where $\mathrm{MAD}_f=\operatorname{median}_i|d_{i,f}-m_f|$. The term $m_f-d_{i,f}$ measures the client's recall deficit on slice $f$. Dividing by $\mathrm{MAD}_f+\epsilon$ accounts for variation across clients, and taking the maximum identifies the feature with the largest normalized deficit.

We use the median and median absolute deviation (MAD) because the cohort may include compromised clients. These statistics tolerate contamination of fewer than half the observations, consistent with our assumption $M<N/2$. We also use the maximum across features to preserve localized anomalous responses that could be diluted by averaging across $|F_{\mathrm{probe}}|$ slices. The one-sided statistic penalizes low recall, while normalization accounts for changes in honest recall during training.

We convert suspicion into a preliminary trust weight using the dead zone $\tau$, gate parameter $\beta$, and clean-root accuracy $a_i$:
\begin{equation}
\tilde t_i=\frac{a_i\,e^{-\beta\max(0,\,s_i-\tau)}}{\sum_j a_j\,e^{-\beta\max(0,\,s_j-\tau)}}.
\label{eq:trust}
\end{equation}
The coordinator measures $a_i$ by evaluating client $i$'s submitted model on $\mathcal{D}_r$. This is the same type of accuracy represented by $\mathrm{Acc}_i$ in Eq.~\eqref{eq:aggregate}, but computed using trusted data at the coordinator.

Because $s_i$ is recomputed every round, some honest clients may have below-median recall even when no attack is present. We apply the penalty only when suspicion exceeds $\tau$. In development runs, this reduced honest false flags from $20.5\%$ to $0.3\%$ without reducing attacker flagging. The exponential gate yields positive preliminary trust when $a_i>0$. We smooth trust across rounds using an exponential moving average (EMA), $t_i(t)=\rho\tilde t_i(t)+(1-\rho)t_i(t-1)$, with $t_i(0)=1/N$. Finally, we aggregate the scaled updates using a coordinate-wise median:
\begin{equation}
\omega(t+1)=\operatorname*{median}_i\Big(\omega(t)+N\,t_i\big(\omega_i^{I}(t)-\omega(t)\big)\Big).
\label{eq:agg}
\end{equation}
The factor $N$ preserves the scale of the updates. With uniform trust, $Nt_i=1$ and Eq.~\eqref{eq:agg} reduces to the coordinate-wise median of the submitted models.

Behavioral trust provides a client-level suspicion score, although its reliability depends on the cohort statistics in Eq.~\eqref{eq:susp}. We examine this limitation in Section~\ref{sec:noniid}. Coordinate-wise median aggregation limits the influence of outlying updates even when the probe-based scores become less informative. The coordinator computes every quantity in Eqs.~\eqref{eq:slice}--\eqref{eq:agg} from its trusted data and the submitted models. Since the proposed rule does not use $\mathrm{Acc}_i$, the accuracy-inflation component A3 has no direct effect on the aggregation weights.

The coordinator performs $|F_{\mathrm{probe}}|N$ forward passes on $\mathcal{D}_r^{\mathrm{spoof}}$ and one root-set accuracy evaluation per client in each round. This cost grows linearly with the number of clients and probed features and requires no backward passes. Client computation and uplink requirements remain unchanged. Algorithm~\ref{alg:round} summarizes the procedure.

\begin{algorithm}[!t]
\caption{One round of receiver-domain behavioral probing}
\label{alg:round}
\begin{algorithmic}[1]
\Require Global model $\omega$; clients $1{:}N$; root set $\mathcal{D}_r$; probes $\{\mathcal{S}_f\}$; $\beta,\tau,\rho$; previous trust $t_i^{\mathrm{prev}}$
\State \emph{Clients (in parallel):} $\omega_i\gets\textsc{LocalTrain}(\omega,\mathcal{D}_i)$; submit $\omega_i$
\State \emph{Coordinator:} $a_i\gets\textsc{Acc}(\omega_i,\mathcal{D}_r)$, $d_{i,f}\gets\textsc{Recall}(\omega_i,\mathcal{S}_f)$ for each $i,f$
\State Compute $m_f$ and $\mathrm{MAD}_f$ from $\{d_{i,f}\}_i$
\State $s_i\gets\max_f\max(0,(m_f-d_{i,f})/(\mathrm{MAD}_f+\epsilon))$
\State Compute and normalize $\tilde t_i$ using~\eqref{eq:trust}
\State $t_i\gets\rho\tilde t_i+(1-\rho)t_i^{\mathrm{prev}}$ (use $t_i^{\mathrm{prev}}=1/N$ initially)
\State $\omega\gets\operatorname{median}_i\bigl(\omega+Nt_i(\omega_i-\omega)\bigr)$
\State \Return $\omega,\{t_i\}$
\end{algorithmic}
\end{algorithm}

\section{Experimental Setup}
\label{sec:setup}

\begin{table*}[!tb]
\caption{Comparison of aggregation methods under identical experimental conditions. $^{\dagger}$ denotes attacker exclusion; other entries report client flagging.}
\label{tab:main}
\centering
\footnotesize
\setlength{\tabcolsep}{7pt}
\renewcommand{\arraystretch}{1.22}
\begin{tabular}{@{}l ccc c cc c@{}}
\toprule
 & \textbf{Clean} & \textbf{Spoofing} & & \textbf{Backdoor} & \textbf{Attacker} & \textbf{Honest} & \textbf{Server} \\
\textbf{Aggregation Rule} & \textbf{Accuracy} & \textbf{Recall} & \textbf{BSR} & \textbf{Lift} & \textbf{Flag/Excl.} & \textbf{False Flag} & \textbf{(ms/Round)} \\
\midrule
Honest FedAvg (No Attack) & $0.7112$ & $0.5292$ & $0.6367$ & $+0.0000$ & --- & --- & $1.0$ \\
\midrule
FedAvg (Attacked) & $0.6932$ & $0.3641$ & $0.8782$ & $+0.2415 \pm 0.0048$ & --- & --- & $1.0$ \\
Accuracy-Weighted FedAvg~\cite{chai2025navigation} & $0.6897$ & $0.3524$ & $0.9402$ & $+0.3036 \pm 0.0124$ & $0\%$ & $0.0\%$ & $1.3$ \\
Coordinate-Wise Median & $0.7101$ & $0.5008$ & $0.7013$ & $+0.0646 \pm 0.0074$ & --- & --- & $0.9$ \\
Trimmed Mean ($f{=}2$) & $0.7061$ & $0.4808$ & $0.7154$ & $+0.0787 \pm 0.0065$ & --- & --- & $1.4$ \\
Krum ($f{=}2$)~\cite{blanchard2017machine} & $0.7071$ & $0.5168$ & $0.6698$ & $+0.0331 \pm 0.0035$ & $100\%^{\dagger}$ & --- & $1.4$ \\
Multi-Krum ($f{=}2$)~\cite{blanchard2017machine} & $0.7108$ & $0.5282$ & $0.6428$ & $+0.0061 \pm 0.0041$ & $100\%^{\dagger}$ & --- & $1.4$ \\
FLTrust~\cite{cao2021fltrust} & $0.6962$ & $0.4687$ & $0.7154$ & $+0.0787 \pm 0.0176$ & $84.7\%$ & $0.0\%$ & $204.4$ \\
\midrule
Behavioral Trust (Ours) & $0.7112$ & $0.5307$ & $0.6406$ & $+0.0039 \pm 0.0047$ & $100\%$ & $0.3\%$ & $38.0$ \\
\;\;$+$ Coordinate-Wise Median (Full) & $0.7143$ & $0.5560$ & $0.6101$ & $-0.0265 \pm 0.0174$ & $100\%$ & $\mathbf{0.3\%}$ & $38.6$ \\
\bottomrule
\end{tabular}
\end{table*}

We use the software-defined-radio GPS spoofing dataset in~\cite{aissou2022dataset} to simulate the UAV clients. Because the dataset contains observations from a single receiver, our experiments evaluate the aggregation mechanisms under simulated client partitions and do not represent a deployed multi-UAV system.

After deduplication and removal of three identifier columns, ten signal-quality features remain. Eight meet the threshold $d\ge0.05$ and form $F_{\mathrm{probe}}$, with $d$ ranging from $0.306$ to $0.152$. The two excluded features have values of $0.018$ and $0.001$, so a trigger in either feature would offer little discriminative information. We subsample $150{,}000$ rows with a $60/40$ authentic/spoofed ratio. The disjoint partitions contain $30{,}000$ test rows, $6{,}000$ root-set rows, and $114{,}000$ training rows distributed across $N=10$ clients, of which $M=2$ are compromised. Each client trains a $10\text{-}64\text{-}32\text{-}16\text{-}1$ MLP ($3{,}329$ parameters) for $3$ local epochs per round over $12$ rounds with Adam at $10^{-3}$ and batch size $512$. We use $p=0.40$, $\gamma=3.0$, and $\widehat{\mathrm{Acc}}_m=0.99$. The default trigger uses $f^\star=\mathrm{CN0}$ and $v_{f^\star}=46.706$~dB-Hz. For the defense, we fix $\beta=1.0$, $\tau=2.0$, and $\rho=0.5$ across all experiments. We keep the data split fixed at seed $42$ and vary client partitioning, poison selection, initialization, and minibatch order over seeds $42$, $7$, and $123$. We report means over these three seeds and pair each seed's lift with its honest-FedAvg baseline.

\section{Results and Discussion}

\subsection{Attack Performance and Attacker-Count Sensitivity}

Table~\ref{tab:main} compares the aggregation methods using the same data split, attack, and random seeds. Krum, Multi-Krum, and trimmed mean use the configured tolerance $f=2$. For Krum and Multi-Krum, the dagger denotes the fraction of malicious client-rounds excluded from selection. The other entries measure low assigned weights, as defined in Section~II.

Under attack, FedAvg yields a backdoor lift of $+0.2415$, and clean accuracy decreases by less than two percentage points. Accuracy-weighted FedAvg with inflated reports produces a larger lift of $+0.3036$. Thus, the accuracy-weighted rule performs worse than FedAvg in this setting. A truthful-report control would be needed to isolate the effect of accuracy inflation. Coordinate-wise median reduces lift to $+0.0646$, while FLTrust achieves $+0.0787$. FLTrust also has lower spoofing recall than the honest baseline. Its norm-based update scaling may reduce the influence of honest updates, although this explanation has not been isolated experimentally. FLTrust assigns attackers a mean trust of $0.039$ and flags them in $84.7\%$ of client-rounds. Its update-direction comparison may be less sensitive to a backdoor affecting a small portion of the $3{,}329$-parameter model.

Multi-Krum also performs well. Its lift of $+0.0061$ is close to that of behavioral trust alone ($+0.0039$), with considerably lower processing cost. Krum ($+0.0331$) and trimmed mean ($+0.0787$) are less effective in this experiment. Behavioral probing offers an additional benefit through client-level flagging, which we discuss in Section~\ref{sec:attrib}.

We next vary the number of compromised clients.

\begin{figure}[!t]
\centering
\includegraphics[width=0.9\columnwidth]{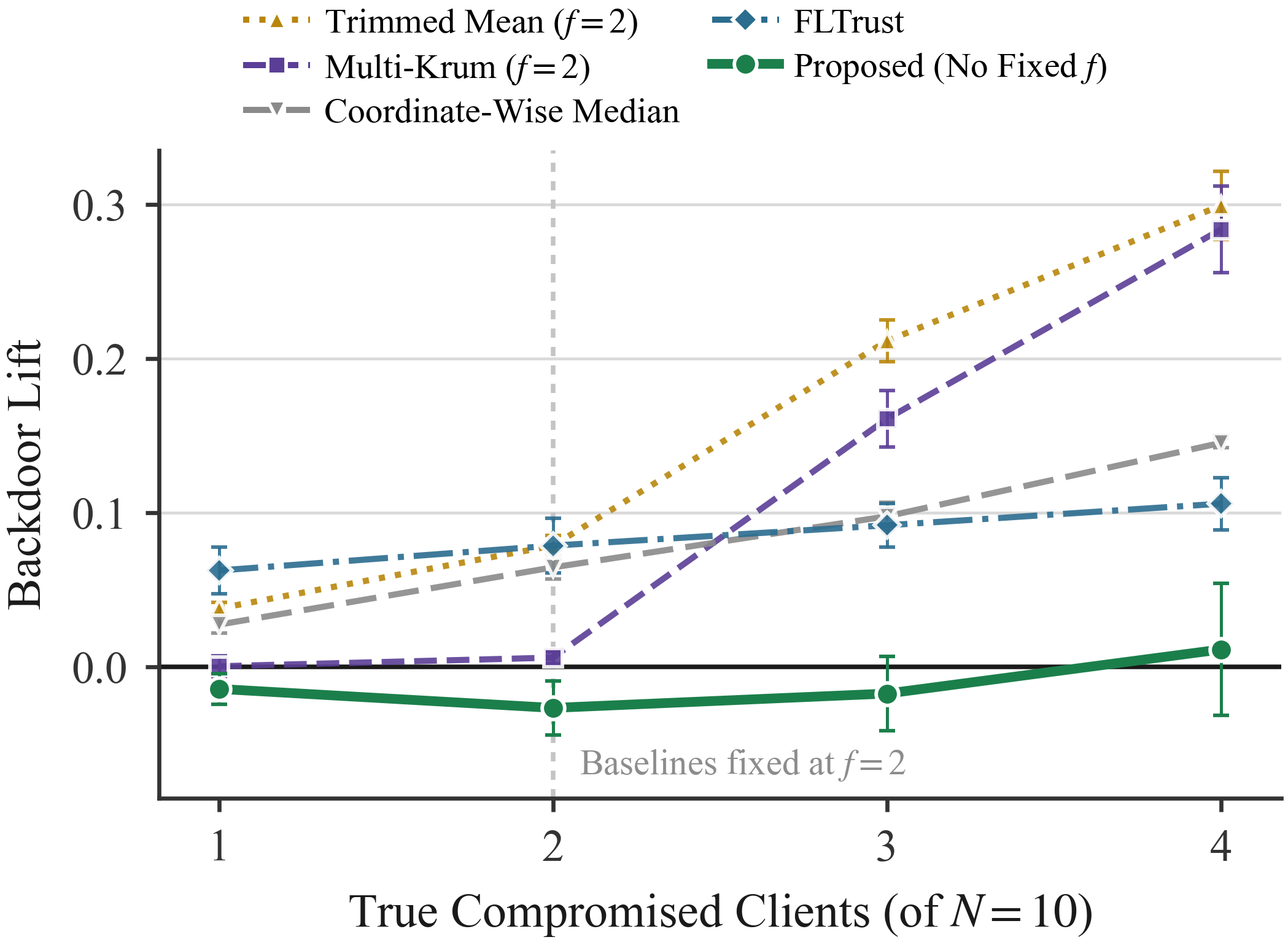}
\caption{Backdoor lift versus the number of compromised clients. Baselines use $f=2$; error bars show standard deviations over three seeds.}
\label{fig:fcount}
\end{figure}

We examine the sensitivity to an underestimated attacker count in Fig.~\ref{fig:fcount}. The Byzantine-robust baselines remain configured for $f=2$, while we vary the actual number of compromised clients from one to four. Multi-Krum has lifts of $+0.0005$ and $+0.0061$ with one and two attackers, respectively. When the actual count exceeds the configured tolerance, its lift rises to $+0.1609$ with three attackers and $+0.2837$ with four, approaching attacked FedAvg. Trimmed mean reaches $+0.2992$. Our method uses no configured tolerance parameter and maintains lift between $-0.0265$ and $+0.0114$. Krum remains relatively stable ($+0.0172$ to $+0.0331$), although it selects only one of the ten updates in each round.

\subsection{Effects of Client Heterogeneity}
\label{sec:noniid}

\begin{figure*}[!tb]
\centering
\includegraphics[width=0.9\textwidth]{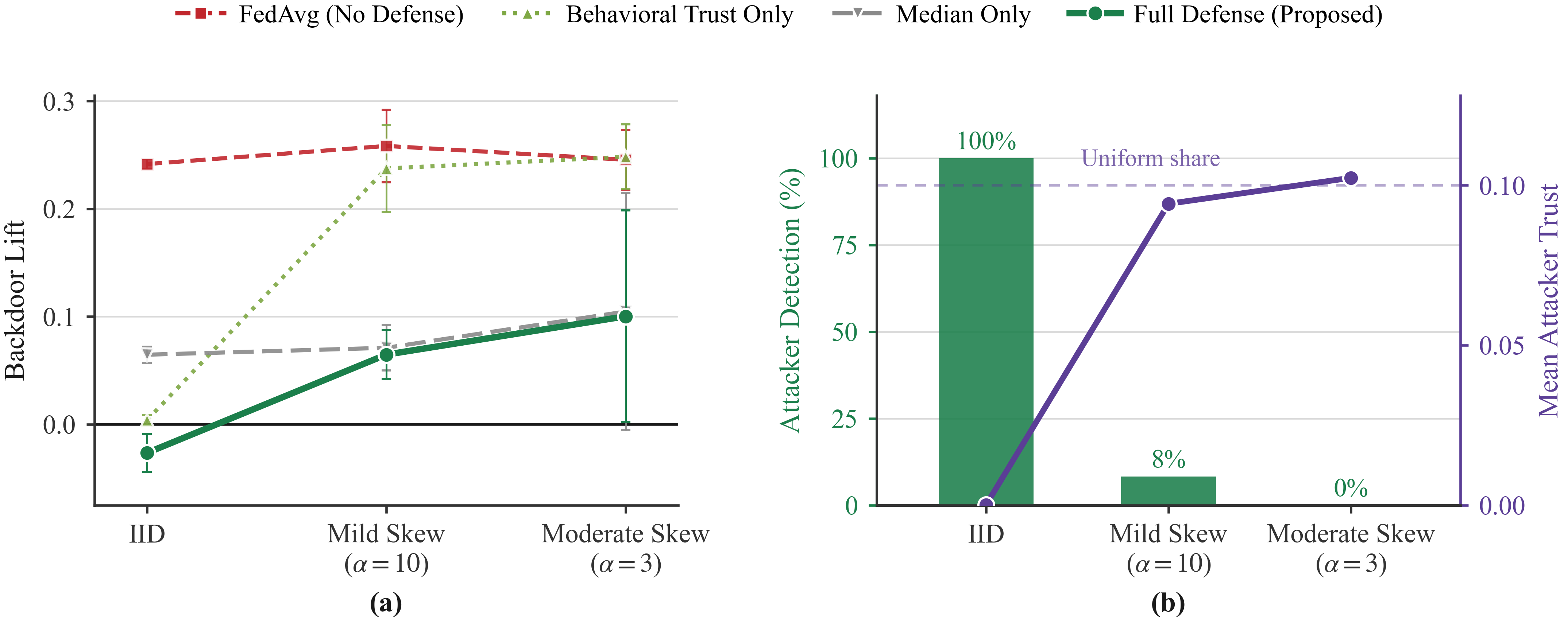}
\caption{Effects of client class-ratio skew: (a) backdoor lift and (b) attacker flagging and mean trust.}
\label{fig:noniid}
\end{figure*}

The suspicion score in Eq.~\eqref{eq:susp} assumes that honest clients respond similarly to the probe slices. To examine this assumption, we create equal-sized client
partitions with different authentic/spoofed class ratios.
The spoofed fraction ranges from $0.19$ to $0.59$ for
$\alpha=10$ and from $0.12$ to $0.72$ for $\alpha=3$,
compared with the global fraction of $0.40$.
We retrain the honest baseline for each setting and
calculate lift against the corresponding reference.
Fig.~\ref{fig:noniid} presents the resulting backdoor
lift, attacker detection rate, and mean attacker trust.

Under IID data, mean attacker trust is $0.0001$, compared with the uniform share of $0.100$, and all compromised client-rounds are flagged. With mild and moderate skew, mean attacker trust rises to $0.0942$ and $0.1023$, and flagging drops to $8\%$ and $0\%$. Trust-only aggregation then gives lifts of $+0.2374$ and $+0.2482$, close to attacked FedAvg on the same partitions. The full defense performs better, with lifts of $+0.0647$ and $+0.1002$. These values are close to coordinate-wise median alone ($+0.0710$ and $+0.1046$), which indicates that the median layer provides most of the protection under skew.

This behavior follows from the normalization in Eq.~\eqref{eq:susp}. Under IID conditions, the client responses are similar and $\mathrm{MAD}_f$ is small, so a backdoored client's deficit spans several MADs. Under class-ratio skew, honest clients respond differently and $\mathrm{MAD}_f$ increases, which reduces the normalized deficit for the same absolute difference. Reducing $\tau$ to $0.5$ improves attacker flagging to only $38.9\%$ at $\alpha=10$ and increases honest false flags to $11.8\%$. This suggests that the main limitation arises from cohort-relative scaling. Per-client normalization may help, while the median layer continues to protect against outlying updates when the probe scores are less informative.
\vspace{-8pt}
\subsection{Trigger Features and Client Attribution}
\label{sec:attrib}

\begin{figure}[!t]
\centering
\includegraphics[width=0.9\columnwidth]{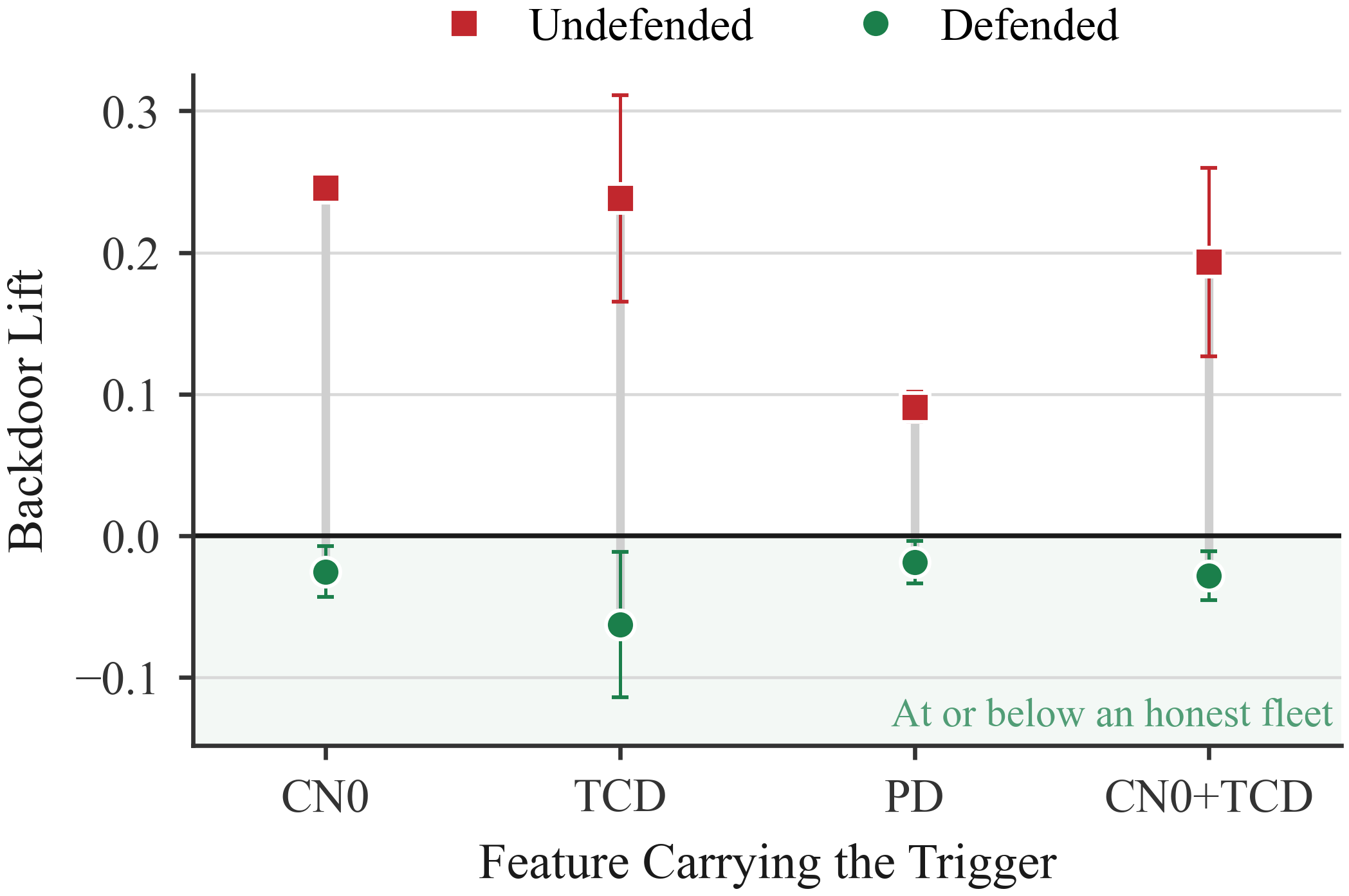}
\caption{Backdoor lift for four trigger settings before and after defense. The shaded region indicates lift at or below the honest baseline.}
\label{fig:trigger}
\end{figure}

We examine four trigger settings in Fig.~\ref{fig:trigger}
without retuning the defense: carrier-to-noise density ratio
(CN0), carrier Doppler in the tracking loop (TCD), pseudorange
(PD), and CN0$+$TCD. All four produce positive backdoor lift
before defense and negative lift afterward. PD is the weakest
probed feature ($d=0.152$). Two additional features, carrier
Doppler (DO) and late-correlator magnitude (LC), give defended
lifts of $-0.0470$ and $-0.0056$, respectively, under the same
configuration. Our coordinator constructs a probe slice for
each discriminative feature, so it does not need to know
which probed feature carries the trigger.

Behavioral probing also provides client-level attribution. Selection rules identify excluded updates, while our method assigns a trust weight to every client. Over $12$ rounds and three seeds, we flag all $72$ compromised client-rounds, whose mean trust is $0.000$. The eight honest clients have mean trust between $0.119$ and $0.131$, compared with the uniform share of $0.100$. Only one of $288$ honest client-rounds is flagged, and none receives zero trust. A false flag therefore reduces a client's influence during a round without excluding it.
\vspace{-3pt}

\subsection{Additional Robustness and Cost Analysis}
\label{sec:cost}

We further examine adaptive evasion, root-set size and corruption, parameter sensitivity, and processing cost. In the tested adaptive attack, an evasion term with weight $\lambda$ encourages compromised models to respond to probes like the honest cohort. At $\lambda=2$, mean attacker trust increases from $0.0001$ to $0.0834$, while undefended lift falls from $+0.2457$ to $-0.1433$; it reaches $-0.4782$ at $\lambda=10$. Defended lift remains between $-0.0253$ and $-0.0829$. These objectives conflict because backdoor training encourages authentic predictions on benign-high triggered inputs, whereas evasion encourages spoofed predictions on the same probes. We have not evaluated other adaptive objectives, including attacks that target the cohort statistics.

Reducing the clean root set tenfold to $600$ rows changes defended lift to $-0.0276$ with $100\%$ attacker flagging. When we flip $30\%$ of its labels, flagging decreases to $83.3\%$, which illustrates the sensitivity of root-of-trust methods, including FLTrust~\cite{cao2021fltrust}, to reference-set corruption. Varying $\beta$ over a sixteen-fold range changes lift by less than one standard deviation, and $\tau$ has a knee at $2.0$. Server processing takes $38.6$~ms per round with ten clients (Table~\ref{tab:main}), compared with $1.0$~ms for FedAvg, and increases to $139.1$~ms with forty clients. We do not claim scalability to large fleets. Probing a subset of clients each round could reduce this cost, although we have not evaluated that option.

\section{Conclusion}

In this paper, we investigated the vulnerability of federated UAV GPS spoofing detection to manipulated client-reported accuracy. With two compromised clients among ten, accuracy-weighted aggregation under poisoned updates and inflated reports produces a backdoor lift of $+0.3036$, compared with $+0.2415$ under attacked FedAvg. We developed receiver-domain behavioral probing, which evaluates submitted models using counterfactual spoofed samples generated at the coordinator. Under IID client data, our method reduces lift to $-0.0265$ and flags compromised client-rounds with a $0.3\%$ honest false-flag rate. It assumes an honest majority but requires no configured Byzantine-tolerance parameter. With strong client heterogeneity, cohort-relative suspicion becomes less reliable and coordinate-wise median aggregation provides most of the protection. The absolute accuracy of the detector also remains limited in the present experiments. We plan to evaluate the method with more heterogeneous UAV fleets and better-converged detectors in future work.

\bibliographystyle{IEEEtran}
\bibliography{ref.bib}

@article{mekdad2023survey,
  author  = {Mekdad, Yassine and Aris, Ahmet and Babun, Leonardo and
             El Fergougui, Abdeslam and Conti, Mauro and
             Lazzeretti, Riccardo and Uluagac, A. Selcuk},
  title   = {A Survey on Security and Privacy Issues of {UAVs}},
  journal = {Computer Networks},
  volume  = {224},
  year    = {2023},
  note    = {Art. no. 109626},
  doi     = {10.1016/j.comnet.2023.109626}
}

@inproceedings{mcmahan2017communication,
  author    = {McMahan, H. B. and Moore, E. and Ramage, D. and
               Hampson, S. and Ag{\"u}era y Arcas, B.},
  title     = {Communication-Efficient Learning of Deep Networks from Decentralized Data},
  booktitle = {Proc. Int. Conf. Artif. Intell. Stat. (AISTATS)},
  pages     = {1273--1282},
  year      = {2017}
}

@article{udin2025federated,
  author  = {Ud Din, Ikram and Taj, Imran and Almogren, Ahmad and Guizani, Mohsen},
  title   = {Federated Learning for Trust Enhancement in {UAV}-Enabled {IoT} Networks: A Unified Approach},
  journal = {IEEE Internet Things J.},
  volume  = {12},
  number  = {8},
  pages   = {9596--9603},
  year    = {2025},
  doi     = {10.1109/JIOT.2025.3542268}
}

@article{chai2025navigation,
  author  = {Chai, Y. and Liu, M. and Li, M.},
  title   = {Navigation Spoofing and Jamming Signals Identification of {UAV} Based on Federated Learning},
  journal = {IEEE Internet Things J.},
  volume  = {12},
  number  = {21},
  pages   = {44177--44188},
  year    = {2025},
  doi     = {10.1109/JIOT.2025.3588162}
}

@article{mowla2020federated,
  author  = {Mowla, N. I. and Tran, N. H. and Doh, I. and Chae, K.},
  title   = {Federated Learning-Based Cognitive Detection of Jamming Attack in Flying Ad-Hoc Network},
  journal = {IEEE Access},
  volume  = {8},
  pages   = {4338--4350},
  year    = {2020}
}

@inproceedings{bagdasaryan2020backdoor,
  author    = {Bagdasaryan, E. and Veit, A. and Hua, Y. and Estrin, D. and Shmatikov, V.},
  title     = {How to Backdoor Federated Learning},
  booktitle = {Proc. Int. Conf. Artif. Intell. Stat. (AISTATS)},
  pages     = {2938--2948},
  year      = {2020}
}

@inproceedings{fang2020local,
  author    = {Fang, M. and Cao, X. and Jia, J. and Gong, N. Z.},
  title     = {Local Model Poisoning Attacks to {Byzantine}-Robust Federated Learning},
  booktitle = {Proc. USENIX Security Symp.},
  pages     = {1605--1622},
  year      = {2020}
}

@inproceedings{blanchard2017machine,
  author    = {Blanchard, P. and El Mhamdi, E. M. and Guerraoui, R. and Stainer, J.},
  title     = {Machine Learning with Adversaries: {Byzantine} Tolerant Gradient Descent},
  booktitle = {Adv. Neural Inf. Process. Syst. (NeurIPS)},
  pages     = {119--129},
  year      = {2017}
}

@inproceedings{cao2021fltrust,
  author    = {Cao, X. and Fang, M. and Liu, J. and Gong, N. Z.},
  title     = {{FLTrust}: {Byzantine}-Robust Federated Learning via Trust Bootstrapping},
  booktitle = {Proc. Netw. Distrib. Syst. Security Symp. (NDSS)},
  year      = {2021}
}

@article{nguyen2024backdoor,
  author  = {Nguyen, Thuy Dung and Nguyen, Tuan and Nguyen, Phi Le and
             Pham, Hieu H. and Doan, Khoa D. and Wong, Kok-Seng},
  title   = {Backdoor Attacks and Defenses in Federated Learning: Survey, Challenges and Future Research Directions},
  journal = {Eng. Appl. Artif. Intell.},
  volume  = {127},
  year    = {2024},
  note    = {Art. no. 107166},
  doi     = {10.1016/j.engappai.2023.107166}
}

@article{mothukuri2021survey,
  author  = {Mothukuri, Viraaji and Parizi, Reza M. and Pouriyeh, Seyedamin and
             Huang, Yan and Dehghantanha, Ali and Srivastava, Gautam},
  title   = {A Survey on Security and Privacy of Federated Learning},
  journal = {Future Gener. Comput. Syst.},
  volume  = {115},
  pages   = {619--640},
  year    = {2021},
  doi     = {10.1016/j.future.2020.10.007}
}

@misc{aissou2022dataset,
  author       = {Aissou, Ghilas},
  title        = {A Dataset for {GPS} Spoofing Detection on Unmanned Aerial System},
  howpublished = {Mendeley Data},
  note         = {Version 3},
  year         = {2022},
  doi          = {10.17632/z7dj3yyzt8.3},
  url          = {https://data.mendeley.com/datasets/z7dj3yyzt8/3}
}

@inproceedings{yin2018byzantine,
  author    = {Yin, D. and Chen, Y. and Ramchandran, K. and Bartlett, P.},
  title     = {{Byzantine}-Robust Distributed Learning: Towards Optimal Statistical Rates},
  booktitle = {Proc. Int. Conf. Mach. Learn. (ICML)},
  pages     = {5650--5659},
  year      = {2018}
}

@article{kang2019incentive,
  author  = {Kang, J. and Xiong, Z. and Niyato, D. and Xie, S. and Zhang, J.},
  title   = {Incentive Mechanism for Reliable Federated Learning: A Joint Optimization Approach to Combining Reputation and Contract Theory},
  journal = {IEEE Internet Things J.},
  volume  = {6},
  number  = {6},
  pages   = {10700--10714},
  year    = {2019}
}

@inproceedings{Xue2023OverTheAir,
  author    = {X. Xue and M. K. Hasan and S. Yu and L. N. Kandel and M. Song},
  title     = {Over-the-Air Federated Learning with Enhanced Privacy},
  booktitle = {Proc. IEEE Int. Conf. Commun. (ICC)},
  year      = {2023},
  address   = {Rome, Italy},
  pages     = {4546--4551}
}

@article{kairouz2021advances,
  author  = {Kairouz, Peter and others},
  title   = {Advances and Open Problems in Federated Learning},
  journal = {Found. Trends Mach. Learn.},
  volume  = {14},
  number  = {1--2},
  pages   = {1--210},
  year    = {2021},
  doi     = {10.1561/2200000083}
}

\end{document}